\documentclass[aps,prl,twocolumn,superscriptaddress]{revtex4-1}

\usepackage{blindtext}
\usepackage{amsmath,amssymb,mathrsfs}
\usepackage{natbib}
\usepackage{subfigure}
\usepackage{tabularx}
\usepackage{epsfig}
\usepackage{longtable}
\usepackage{amsfonts}
\usepackage{rotating}
\usepackage{subfigure}
\usepackage{amsmath}
\usepackage{comment}
\usepackage{bbold}
\usepackage{color}
\usepackage{braket}

\usepackage[unicode=true,bookmarks=true,bookmarksnumbered=false,bookmarksopen=false,breaklinks=false,pdfborder={0 0 1},backref=false,colorlinks=true]{hyperref}

\hypersetup{linkcolor=magenta,urlcolor=blue,citecolor=blue,pdfstartview={FitH},hyperfootnotes=false,unicode=true}

\def\be{\begin{equation}}
\def\ee{\end{equation}}
\def\bea{\begin{eqnarray}}
\def\eea{\end{eqnarray}}

\begin{document}
\title{Emergent Orbital Skyrmion Lattice in a Triangular Atom Array}
\author{Rui Cao}
\affiliation{Department of Physics, National University of Defense Technology, Changsha 410073, P. R. China}
\author{Jinsen Han}
\affiliation{Department of Physics, National University of Defense Technology, Changsha 410073, P. R. China}
\author{Jianmin Yuan}
\affiliation{Department of Physics, Graduate School of China Academy of Engineering Physics, Beijing 100193, P. R. China}
\affiliation{Department of Physics, National University of Defense Technology, Changsha 410073, P. R. China}
\author{Xiaopeng Li}
\email{xiaopeng\_li@fudan.edu.cn}
\affiliation{State Key Laboratory of Surface Physics, Key Laboratory of Micro and Nano Photonic Structures (MOE), and Department of Physics, Fudan University, Shanghai 200433, China}
\affiliation{Institute for Nanoelectronic Devices and Quantum Computing, Fudan University, Shanghai 200433, China}
\affiliation{Shanghai Qi Zhi Institute, AI Tower, Xuhui District, Shanghai 200232, China}
\author{Yongqiang Li}
\email{li\_yq@nudt.edu.cn}
\affiliation{Department of Physics, National University of Defense Technology, Changsha 410073, P. R. China}

\begin{abstract}
Multi-orbital optical lattices have been attracting rapidly growing research interests in the last several years, providing fascinating opportunities for orbital-based quantum simulations.
Here, we consider bosonic atoms loaded in the degenerate $p$-orbital bands of a two-dimensional triangular optical lattice. This system is described by a multi-orbital Bose-Hubbard model.
We find the confined atoms in this system develop spontaneous orbital polarization, which forms a chiral Skyrmion lattice pattern in a large regime of the phase diagram. This is in contrast to its spin analogue which largely requires spin-orbit couplings. The emergence of the Skyrmion lattice is confirmed in both bosonic dynamical mean-field theory (BDMFT) and exact diagonalization (ED) calculations. By analyzing the quantum tunneling induced orbital-exchange interaction in the strong interaction limit, we find the Skyrmion lattice state arises due to the interplay of $p$-orbital symmetry and the geometric frustration of the triangular lattice. We provide experimental consequences of the orbital Skyrmion state, that can be readily tested in cold atom experiments. Our study implies orbital-based quantum simulations could bring exotic scenarios unexpected from their spin analogue.

\end{abstract}

\date{\today}


\maketitle

{\it Introduction}.
Last several years have witnessed rapid progress in preparing atomic multi-orbital superfluids in optical lattices~\cite{PhysRevLett.121.265301,p-band_honecomb,2021_Zhou_OrbitalControl,p-topological_2021,song2022realizing,hartke2022quantum,PhysRevLett.127.143401,PhysRevLett.127.033201,venu2022observation,PhysRevResearch.4.043083}. Versatile quantum many-body phenomena have been observed by combining multi-orbital setting and complex lattice structure~\cite{p-band_honecomb,p-topological_2021}.
Excited band condensate in a hexagonal lattice has been achieved via a lattice swap technique~\cite{p-band_honecomb}, where a Potts-nematic superfluid appears due to interaction induced quantum fluctuations. Further cooling of this atomic condensate system~\cite{p-topological_2021} leads to a chiral condensate for weakly interacting bosons~\cite{Hemmerich11,PhysRevA.74.013607,PhysRevLett.97.110405,2008_Lim_PRL,2012_Liu_PRL}.
Universal single-qubit control in the $s$- and $d$-orbital subspaces has been implemented with topologically protected robustness~\cite{2021_Zhou_OrbitalControl}. These recent developments open up unprecedented opportunities for orbital-based quantum simulations, by which the realizable quantum many-body states and phenomena could strongly deviate from the spin analogue due to the fundamental difference in their symmetry~\cite{Li_2016}.

\begin{figure}[th!]
\includegraphics[trim = 0mm 0mm 0mm 0mm, clip=true, width=0.45\textwidth]{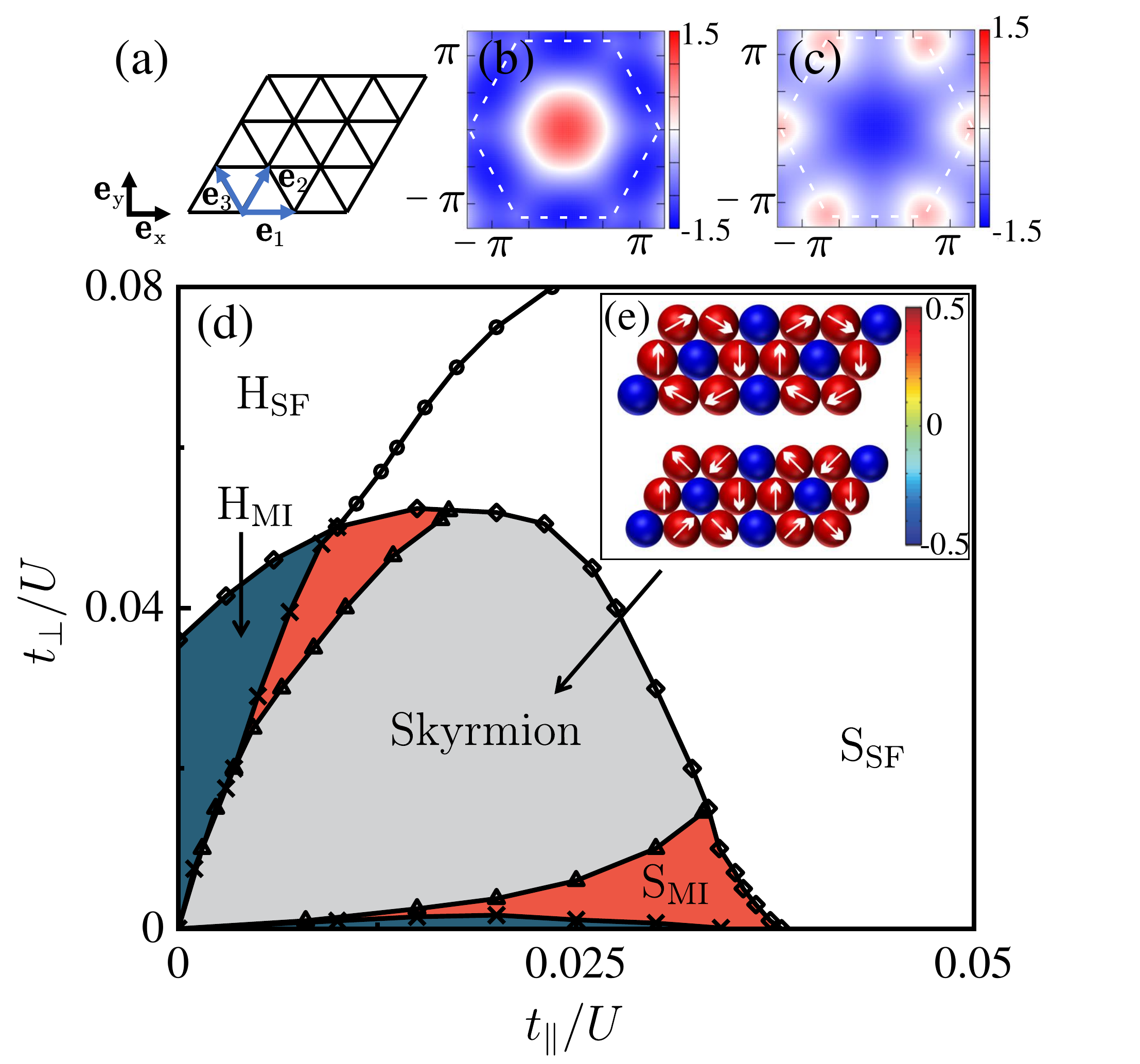}
\vspace{-3mm}
\caption{(Color online) (a) The geometry of the two-dimensional triangular lattice. (b),(c) The energy dispersion of the lowest $p$-orbital band  of  a two-dimensional triangular lattice, with (b) $t_{\perp}=0$, and (c) $t_{\parallel}=0$. (d) Many-body phase diagram in terms of hopping amplitudes $t_\parallel$ and $t_\perp$ for filling $n_{\bf r}=1$, obtained from bosonic dynamical mean-field theory, where the system favors three Mott phases with Skyrmion, stripe (${\rm S_{MI}}$), ferro-orbital (${\rm H_{MI}}$) textures, and two superfluid phases with stripe (${\rm S_{SF}}$) and ferro-orbital (${\rm H_{SF}}$) angular momentum. (e) Real-space orbital textures for Skyrmion (upper) and anti-Skyrmion (down) lattices, where the arrows represent the projection in the $xy$-plane of orbital polarization texture $\langle\boldsymbol{\mathcal S}_{{\bf r}} \rangle$, and the color denotes the $z$ component. The interactions are $U=3U_1=3U_2$.}
\label{fig_combine}
\end{figure}

In this letter, we study interacting $p$-orbital bosons in a two-dimensional triangular optical lattice, which corresponds to the experimental setups of hexagonal lattices~\cite{p-band_honecomb,p-topological_2021} with a large sublattice potential imbalance~\cite{Xu2022}. This system is described by a multi-orbital Bose-Hubbard model~\cite{2009Unconventional,Li_2016}.
The quantum many-body phases of this system are investigated in the strong interaction regime by BDMFT and ED calculations. We find an orbital Skyrmion lattice state emerges in the Mott insulating regime. The Skyrmion lattice state has a composite chirality that spontaneously breaks the time-reversal symmetry. In the orbital setting, two different types of Skyrmion lattices occur simultaneously in the degenerate quantum many-body ground states, which is in sharp contrast to the Dzyaloshinskii-Moriya scenario commonly adopted to create Skyrmion textures in spin systems~\cite{R2006Spontaneous,DZYALOSHINSKY1958241,PhysRev.120.91,09Skyrmion,2010Real}. Near the Mott-superfluid transition, we find the chiral Skyrmion lattice melts to quantum states with stripe and ferro-orbital orders.

{\it Model and Hamiltonian}.
The system of interacting spinless bosonic atoms, loaded into the $p$-orbital bands of a two-dimensional (2D) triangular optical lattice, can be described by a multi-orbital Bose-Hubbard model in the tight-binding limit~\cite{2009Unconventional,Li_2016}
\begin{eqnarray}
\label{eq:Ham}
H &=&t_{\parallel} \sum_{m, {\bf r} }  p_{m, {\bf r}} ^\dag  p_{m,{\bf r} + {\bf e}_m}- t_{\perp}\sum_{m, {\bf r}}  p^{^\prime\dagger}_{m, {\bf r}}  p^\prime_{m,{\bf r} + {\bf e}_m}
+ {\rm H.c.}  \nonumber \\
&+& \frac{U}{2} \sum_{\bf r} n_{\bf r } (n_{\bf r} -1) + 2{U_1}  \sum_{\bf r} n_{x,{\bf r}} n_{y,{\bf r}} \nonumber\\
&+& \frac{U_2}{2}\sum_{\bf r,\nu\neq\nu^\prime} p^\dagger_{\nu,{\bf r}}p^\dagger_{\nu,{\bf r}} p_{\nu^\prime,{\bf r}}p_{\nu^\prime,{\bf r}}-\mu \sum_{\bf r} n_{\bf r},
\end{eqnarray}
where $t_{\parallel}$ and $t_{\perp}$ denote the hopping amplitudes between two nearest-neighboring $p$-orbitals along the parallel and the perpendicular directions, respectively.
The lattice annihilation operators $p_{m,{\bf r}} \equiv (p_{{x},{\bf r}}{\bf e}_x  + p_{{y},{\bf r}}{\bf e}_y)\cdot {\bf e}_m$ with the unit vectors ${\bf e}_1= {\bf e}_x$ and ${\bf e}_{2,3}=\pm \frac{1}{2}{\bf e}_x + \frac{\sqrt{3}}{2}{\bf e}_y$ for hopping $t_{\parallel}$, and $p^\prime_{m,{\bf r}} \equiv ( p_{{x},{\bf r}}{\bf e}_x  + p_{y,{\bf r}}{\bf e}_y\big)\cdot {\bf e}^\prime_m$ with ${\bf e}^\prime_1={\bf e}_y$ and ${\bf e}^\prime_{2,3}=-\frac{\sqrt{3}}{2}{\bf e}_x \pm \frac{1}{2} {\bf e}_y$ for hopping $t_{\perp}$. Here, ${\bf e}_m$ is shown in Fig.~\ref{fig_combine}(a), and $p_{x,{\bf r}}$ ($p_{y,{\bf r}}$) denotes the annihilation operator for the $p_x$ ($p_y$) orbital degree of freedom at site ${\bf r}$. $\mu$ is the chemical potential, $n_{\bf r} = \sum_{\nu} p^\dag_{{\nu},{\bf r}} p_{{\nu},{\bf r}}$ with $\nu=x,y$, and $U$, $U_1$ and $U_2$ denote the interaction strengths with $U-2U_1=U_2$ as a result of symmetry analysis for the triangular lattice. In the deep lattice limit, the harmonic approximation of the Wannier function implies $U=3U_1=3U_2$, and consequently the interactions ($U$ and $U_{1,2}$) take a simplified form as $H_{\rm int}= \frac{U}{2}\sum_{\bf r}( n^2_{\bf r} - \frac{1}{3}L^2_{z,\bf r} )$, with the orbital angular momentum
$L_{z, {\bf r}}\equiv -i(p_{x,{\bf r}} ^\dag p_{y,{\bf r}} - p_{y,{\bf r}} ^\dag p_{x,{\bf r}} )$~\cite{PhysRevA.74.013607}.

{\it Weak interaction limit.}
To understand the many-body phenomena, we first discuss the physics in the weakly interacting superfluid regime with $t_{\parallel,\perp} \gg U$ and $U_{1,2}$, where the bosons are expected to condense. For triangular lattices, the Brillouin zone forms the shape of a regular hexagon with the edge length $4\pi/3$, where the lattice constant is set to be the unit of length. The single-particle spectrum of the noninteracting $p$-band bosonic system is shown in Fig.~\ref{fig_combine}(b),(c), where we plot the dispersion of the lowest $p$-orbital band of a 2D triangular lattice. For $t_{\perp}=0$, the system supports three degenerate minima located at ${\bf M}_0=(0,2\pi/\sqrt{3})$ and ${\bf M}_{\pm}=(\pm\pi,\pi/\sqrt{3})$ [Fig.~\ref{fig_combine}(b)]. For $t_{\parallel}=0$, band minima move to the center of Brillouin zone [Fig.~\ref{fig_combine}(c)]. Due to the competition between hopping $t_\parallel$ and $t_\perp$, it is expected to develop rich orbital orders with the appearance of orbital angular momentum $\langle L_{z,{\bf r}}\rangle \neq 0$. We find that the system supports two different types of condensates in the weakly interacting regime. The ground state in the limit of $t_\perp \ll t_\parallel$ is a stripe superfluid phase [${\rm S_{SF}}$ phase in Fig.~\ref{fig_combine}(d)], which is described by
\bea
\Phi^1_N \propto  [\sum_{\bf r}  {\rm e}^{i({\bf k } \cdot {\bf r} + \beta_{\bf r})}\big({\rm cos}\alpha p^\dagger_{x,{\bf r}} \pm i \sigma_{\bf r} {\rm sin}\alpha p^\dagger_{y,{\bf r}} \big)]^N|0\rangle.
 \eea
In the regime of $t_\parallel \ll t_\perp$, the system condenses at the center of Brillouin zone and demonstrates a ferro-orbital order [${\rm H_{SF}}$ phase in Fig.~\ref{fig_combine}(d)], with
\bea
\Phi^2_N \propto [\sum_{\bf r} {\rm e}^{i{\bf k}\cdot {\bf r}} \big(p^\dagger_{x,{\bf r}} \pm i p^\dagger_{y,{\bf r}} \big)]^N|0\rangle.
\eea
Here, $|0\rangle$ is the vacuum state, $N$ denotes the total number of particles, $\sigma_{\bf r}= \pm 1$ is the sign of the staggered orbital angular momentum, $\beta_{\bf r}= 0$ ($\sigma_{\bf r}=1$) or $\pi/2$ ($\sigma_{\bf r}=-1$) in the stripe direction, and $\beta_{\bf r}=\pi$ in the homogeneous direction. Note here that $\alpha=\pi/6$ in the regime of $t_\parallel \gg t_\perp$ and $t_{\parallel,\perp}/U\gg 1$~\cite{PhysRevLett.97.190406} and the bosons condense at two of the three degenerate minima as observed in the experiments~\cite{Xu2022}, which is consistent with our numerical simulations.

{\it Orbital Skyrmion lattice state at strong interaction.}
Considering the tunability of Hubbard parameters experimentally~\cite{Hemmerich11}, we extend our study to the strongly interacting regime of the spinless $p$-orbital bosons in the triangular lattice, described by Eq.~(\ref{eq:Ham}). To analyze quantum ground states of the many-body system, a bosonic version of dynamical mean-field theory is implemented. We remark here that BDMFT is an extension of fermionic dynamical mean-field theory, and suitable to treat strongly correlated systems for the full range of couplings from Mott insulator to superfluid. To accommodate long-range orders that spontaneously break lattice-translational symmetry, we generalize a real-space BDMFT~\cite{Li2011} for our system of spinless $p$-orbital bosons in the triangular lattice. The technical details can be found in Supplementary Materials~\cite{SM}.

At strong interactions, bosons form a Mott insulating state. Through BDMFT calculation, we find this Mott state develops spontaneous orbital polarization forming a Skyrmion lattice that breaks time-reversal and lattice-translational symmetries.
Interestingly, the BDMFT calculation reveals two different types of Skyrmion textures, $\rm i.e.$, Skyrmion and anti-Skyrmion lattices as shown in Fig.~\ref{fig_combine}(e), where an orbital polarization vector $\langle\boldsymbol{\mathcal S}_{{\bf r}} \rangle = \big(\langle {\mathcal S}^x_{\bf r}  \rangle,
\langle {\mathcal S}^y_{\bf r} \rangle,  \langle{\mathcal S}^z_{\bf r} \rangle \big)$ is defined, with ${\mathcal S}^x_{\bf r} \equiv \frac{1}{2} ({p}_{x,{\bf r}} ^\dagger {p}_{x, {\bf r}} - {p}_{y,{\bf r}} ^\dagger {p}_{y, {\bf r}})$, ${\mathcal S}^y_{ \bf r} \equiv \frac{1}{2}({p}_{x,{\bf r}} ^\dagger {p}_{y , {\bf r}} + {p}_{y,{\bf r}} ^\dagger {p}_{x , {\bf r}})$, and ${\mathcal S}^z_{\bf r} \equiv \frac{1}{2i}({p}_{x,{\bf r}} ^\dagger {p}_{y, {\bf r}} - {p}_{y,{\bf r}} ^\dagger {p}_{x, {\bf r}})$.
The two Skyrmion lattice phases are connected by the $\mathcal{T}\times \mathcal{I}$ symmetry,
with $\mathcal{T}$ and $\mathcal{I}$ being time-reversal and space-reflection ($p_{x,{\bf r}} \rightarrow -p_{x,{\bf r}}$) symmetries, respectively.
We remark here that the emergent orbital Skyrmion texture observed here is solely induced by onsite interactions, and the underlying physics is the interplay of $p$-orbital symmetry and geometric frustration of the triangular lattice, captured by an effective orbital-exchange model (this model will be discussed later).

To show the robustness of Skyrmion texture against quantum fluctuations, we map out the full $t_\parallel-t_\perp$ phase diagram for filling $n_{\bf r}=1$ and interactions $U=3U_1=3U_2$ in the framework of BDMFT, as shown in Fig.~\ref{fig_combine}(d). We find that the orbital Skyrmion lattice is robust against quantum fluctuations and explores a wide regime in Mott phases. Only for sufficiently large asymmetry between the two hopping amplitudes, two other Mott phases develop instead, including stripe-orbital phase (${\rm S_{ MI}}$) breaking time-reversal symmetry, and ferro-orbital phase (${\rm H_{MI}}$) respecting time-reversal symmetry (real-space orbital textures shown in Fig. S1~\cite{SM}). All of these orbital-ordered phases are found to persist up to the superfluid transition. After the Mott-superfluid transition, the system demonstrates two superfluid phases, where one is a stripe-orbital phase (${\rm S_{SF}}$) breaking time-reversal, lattice-translational and rotational symmetries~\cite{PhysRevLett.97.190406}, and the other a ferro-orbital phase (${\rm H_{SF}}$) breaking time-reversal symmetry, consistent with Eq.~(2) and (3). Note here that the phase diagram is symmetric upon orbital interchange in the low-hopping regime, which is also manifested in the effective orbital-exchange model [Eq.~(\ref{eq:eff})].

\begin{figure}
\includegraphics[trim = 0mm 0mm 0mm 0mm, clip=true, width=0.485\textwidth]{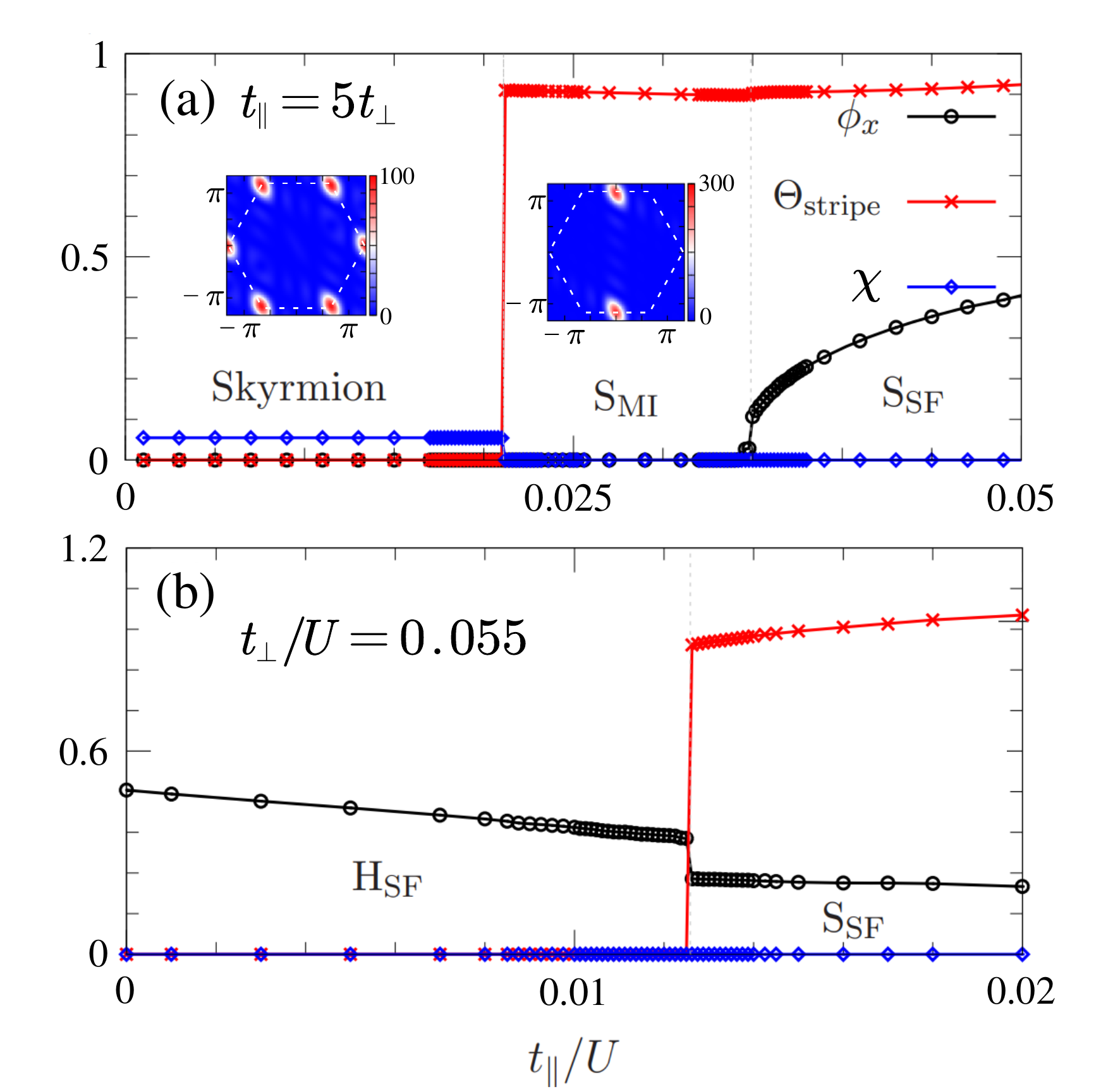}
\caption{(Color online) Phase transitions of ultracold bosonic gases in $p$-orbital bands of a 2D triangular lattice for different hopping amplitudes (a) $t_\parallel=5t_\perp$ and (b) $t_\perp=0.055U$, obtained via bosonic dynamical mean-field theory~\cite{SM}. Indicated by the dashed lines, the system demonstrates (a) a Mott transition from the Skyrmion to the stripe phase (${\rm S_{MI}}$), followed by the appearance of a stripe superfluid phase (${\rm S_{SF}}$), and (b) a superfluid transition from the ferro- (${\rm H_{SF}}$) to the stripe-orbital phase (${\rm S_{SF}}$) upon increasing hopping $t_\parallel$.  Inset: Contour plots of spin structure factor $\mathcal{F}^{z}(\boldsymbol{k})$ for Mott phases. The interactions $U=3U_1=3U_2$, and filling $n_{\bf r}=1$.}
\label{transition_new}
\end{figure}

To quantify phase boundaries in Fig.~\ref{fig_combine}(d), we introduce superfluid order $\phi_{\nu}\equiv\sum_{\bf r}|\langle p_{\nu,{\bf r}}\rangle|/N_{\rm lat}$, stripe order $\Theta_{\rm stripe}\equiv \langle{\mathcal S}^z_{+}\rangle-\langle{\mathcal S}^z_{-}\rangle$, and scalar spin chirality $\chi=\langle \boldsymbol{\mathcal S}_{\boldsymbol{r}} \cdot \left( \boldsymbol{\mathcal S}_{\boldsymbol{r}+{\rm \boldsymbol{e}_1}} \times \boldsymbol{\mathcal S}_{\boldsymbol{r}+{\rm \boldsymbol{e}_2}}\right) \rangle$, where $N_{\rm lat}$ is the number of lattice sites, and $\langle{\mathcal S}^z_{+}\rangle$ ($\langle{\mathcal S}^z_{-}\rangle$) denotes the $z$-component of orbital polarization per site on the stripe with positive (negative) value.
We clearly observe nonzero values of scalar spin chirality $\chi \neq0$ in the Skyrmion phase, as shown in Fig.~\ref{transition_new}(a). For larger hopping amplitudes, we find a Mott phase transition from the Skyrmion to the ${\rm S_{MI}}$ phase, indicated by the absence of $\chi=0$ and the appearance of $\Theta_{\rm stripe}\neq0$. The corresponding contour plots of static spin structure factor ${\mathcal F}^{z}(\boldsymbol{k})=\sum_{i,j}e^{i\boldsymbol{k}\cdot({\bf r}_i-{\bf r}_j)} \langle {\mathcal S}^{z}_{{\bf r}_i} {\mathcal S}^{z}_{{\bf r}_j}\rangle$ are shown in the inset of Fig.~\ref{transition_new}(a) for different Mott phases, where $i$ and $j$ denote the lattice sites. Increasing hopping amplitudes further, atoms delocalize with the coexistence of stripe order $\Theta_{\rm stripe}\neq0$ and superfluid order $\phi_{x,y}\neq0$. In addition, we observe a superfluid phase transition from a ferro- to a stripe-orbital order, as shown in Fig.~\ref{transition_new}(b). We remark here that the phase transitions are found to be discontinues within BDMFT.

{\it Orbital exchange and effective model construction.}
To explain the underlying mechanism of the orbital textures in the deep Mott regime with $t_{\parallel,\perp} \ll U$ and unit filling, we construct an effective orbital-exchange model for Eq.~(\ref{eq:Ham}). In our case, the orbital-exchange interactions arise from the virtual hopping processes induced by $t_{\parallel,\perp}$, and are obtained from the perturbative expansion of tunneling processes up to third order (third-order expansion will be justified later). By introducing the projection operator $\mathcal{P}$ to describe the Hilbert space of the singly occupied Mott state, the effective Hamiltonian reads $H_{\rm eff}\mathcal{P} | \psi \rangle = E \mathcal{P} | \psi \rangle $, where $H_{\rm eff}=-\mathcal{P} H_t\mathcal{Q} \left( \mathcal{Q} H\mathcal{Q} -E \right)^{-1} \mathcal{Q} H_t\mathcal{P} $ with $\mathcal{Q} =1-\mathcal{P} $, and $H_t$ being the hopping part of Eq.~(\ref{eq:Ham}). Due to $E \sim t^2/U$, we obtain $\mathcal{Q} H\mathcal{Q} -E \approx \mathcal{Q} H\mathcal{Q} $.

Generally, the orbital polarization operator $\boldsymbol{\mathcal{S}}_{\bf r}$ changes with bond orientations~\cite{PhysRevLett.97.190406,PhysRevLett.100.160403,PhysRevLett.100.200406,PhysRevB.103.205144}. It is convenient to introduce the rotation direction ${\bf e}_{\theta}$ for orbital polarization operator, {\rm i.e.}, ${\bf e}_{\theta}^x={\rm cos}\left(2 {\rm \theta} \right){\bf{e}}_x+{\rm sin}\left(2{\rm \theta}\right){\bf{e}}_y$, ${\bf e}_{\theta}^y=- {\rm sin}\left(2 {\rm \theta} \right){\bf{e}}_x+{\rm cos}\left(2{\rm \theta}\right){\bf{e}}_y$ and ${\bf e}_{\theta}^z={\bf e}_z$, for a bond directing at angle $\theta$ with the $x$ axis. With the definition above, we finally obtain an anisotropic orbital-exchange model for the triangular lattice system with interactions $U=3U_1=3U_2$~\cite{SM},
\begin{eqnarray}
\label{eq:eff}
H_{\rm eff} &=&\sum_{{\bf r},m,v}\left(J_v+J^{\prime}_v\right) \left[\boldsymbol{\mathcal{S}}_{{\bf r}}\cdot {\bf e}^v_{\theta_m}\right]\left[\boldsymbol{\mathcal{S}}_{{\bf r}+{\bf{e}}_m}\cdot {\bf e}^v_{\theta_m}\right] \\
&+&\sum_{{\bf r},u,v,w}J^{\prime}_{uvw}\left[\boldsymbol{\mathcal{S}}_{\bf r}\cdot {\bf e}^u_{\theta_1}\right]\left[ \boldsymbol{\mathcal{S}}_{{\bf r}+{\bf e}_1} \cdot {\bf e}^v_{\theta_1}\right]\left[\boldsymbol{\mathcal{S}}_{{\bf r}+{\bf e}_2}\cdot {\bf e}^w_{\theta_1}\right],\nonumber
\end{eqnarray}
where $\{u,v,w\}=\{x,y,z\}$, ${\theta_m}$ is the angle with the $x$ axis for the bond ${\bf{e}_m}$, and $J$ and $J^{\prime}$ denote the orbital-exchange terms from second- $\mathcal{O}(t^2_{\parallel,\perp}/U)$ and third-order $\mathcal{O}(t^3_{\parallel,\perp}/U^2)$ tunneling processes, respectively. In the absence of third-order interactions, the effective model reduces to an $XYZ$ model with orbital-exchange parameters $J_x=-3(t^2_\parallel+t^2_\perp)/2U$, $J_y=3t_\parallel t_\perp/U$, and $J_z=9t_\parallel t_\perp/U$~\cite{Li_2016}. Generally, $J_x$ dominates in the regime of $t_\perp \ll t_\parallel$ or $t_\parallel \ll t_\perp$, where in-plane ferro-orbital order develops, and $J_z$ dominates the remains ($t_\parallel \approx t_\perp$), where the system favors out-of-plane Ising-orbital order for bipartite lattices. For triangular lattices, however, the exchange coupling $J_z$ ($t_\parallel \approx t_\perp$) results in Ising-type frustration~\cite{PhysRev.79.357,1977Toulouse,moessner2006geometrical, nisoli2013colloquium,PhysRevA.86.043620} forming novel quantum phases~\cite{balents2010spin,nisoli2013colloquium}. In addition, the orbital-exchange interactions in Eq.~(\ref{eq:eff}) are strongly anisotropic, as a result of the anisotropic $p$-orbital hopping, leading to unique properties in the triangular lattice, as shown below.
\begin{figure}
\includegraphics[trim = 0mm 0mm 0mm 0mm, clip=true, width=0.5\textwidth]{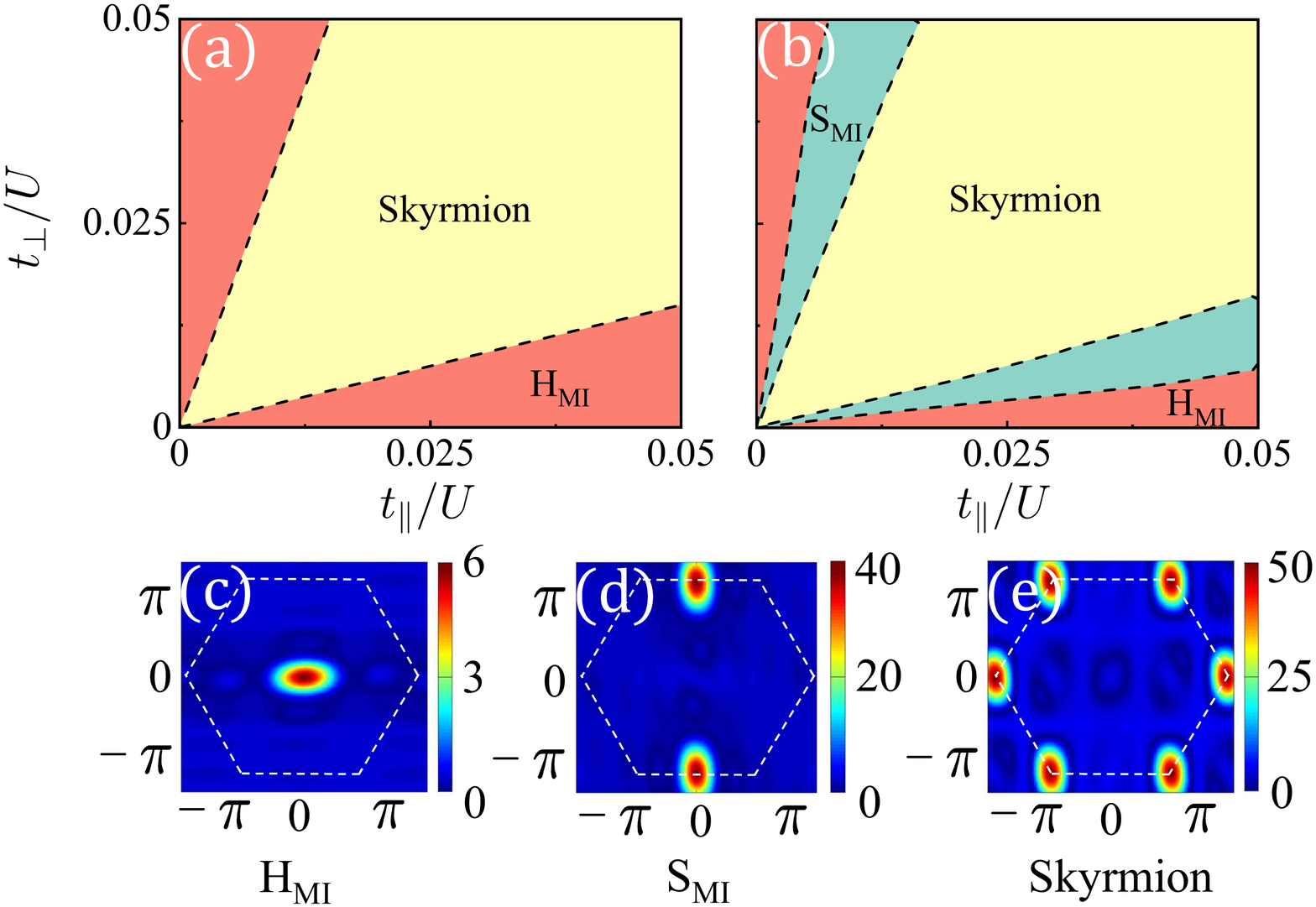}
\caption{(Color online) Phase diagrams of (a) second- and (b) third-order anisotropic orbital-exchange models as a function of hopping amplitudes $t_\parallel$ and $t_\perp$, obtained from exact diagonalizations for a lattice of $N_{\rm lat}=24$ sites~\cite{SM}. The effective model with third-order exchange interactions favors three Mott-insulating phases with Skyrmion, stripe ($\rm S_{MI}$), and ferro-orbital ($\rm H_{MI}$) textures, consistent with BDMFT results. (c-e) Contour plots of spin structure factor ${\mathcal F}^{z}(\boldsymbol{k})$ for different Mott phases. The interactions are $U=3U_1=3U_2$.}
\label{ed_phase}
\end{figure}

We numerically solve the frustrated orbital-exchange model by ED with Quspin python package~\cite{weinberg2017quspin,weinberg2019quspin}. Here, we mainly consider lattices with periodic boundary conditions ~\cite{SM}. Phase diagrams of the orbital-exchange model are shown in Fig.~\ref{ed_phase}(a),(b). To distinguish different Mott-insulating phases, both fidelity metric $g$ (Fig. S4~\cite{SM})~\cite{gu2010fidelity,PhysRevE.74.031123,PhysRevB.82.115125} and static spin structure factor ${\mathcal F}^{z}(\boldsymbol{k})$ are utilized. Within ED, we find that the orbital Skyrmion phase is described by the effective orbital-exchange model with leading-order $\mathcal{O}(t^2_{\parallel,\perp})/U^2$ tunneling processes, as shown in Fig.~\ref{ed_phase}(a). Considering the absence of Dzyaloshinsky-Moriya interactions~\cite{R2006Spontaneous,DZYALOSHINSKY1958241,PhysRev.120.91,09Skyrmion,2010Real} in the effective model, the mechanism for generating orbital Skyrmion texture is a result of the interplay of hopping-induced anisotropic orbital-exchange interactions and geometric
frustration of the triangular lattice. The leading-order orbital-exchange model, however, only favors two Mott phases, i.e., Skyrmion and $\rm H_{MI}$ [Fig.~\ref{ed_phase}(a)].
After including subleading-order $\mathcal{O}(t^3_{\parallel,\perp}/U^2)$ tunneling processes, ED resolves three Mott phases, i.e., Skyrmion, $\rm S_{MI}$ and $\rm H_{MI}$ [Fig.~\ref{ed_phase}(b)], whose conclusion is consistent with the prediction of BDMFT for the extended Bose-Hubbard model [Fig.~\ref{fig_combine}(d)]. The corresponding static spin structure factors ${\mathcal F}^{z}(\boldsymbol{k})$ for each phase are shown in Fig.~\ref{ed_phase}(c-e) (consistent with the inset of Fig.~\ref{transition_new}(a), obtained from BDMFT). We remark here that subleading-order orbital-exchange interactions are absent in the previous studies~\cite{PhysRevLett.97.190406,PhysRevLett.100.160403,PhysRevLett.100.200406,PhysRevB.103.205144} and should be included in the effective model to obtain the complete Mott phases.

\begin{figure}
\includegraphics[trim = 0mm 0mm 0mm 0mm, clip=true, width=0.475\textwidth]{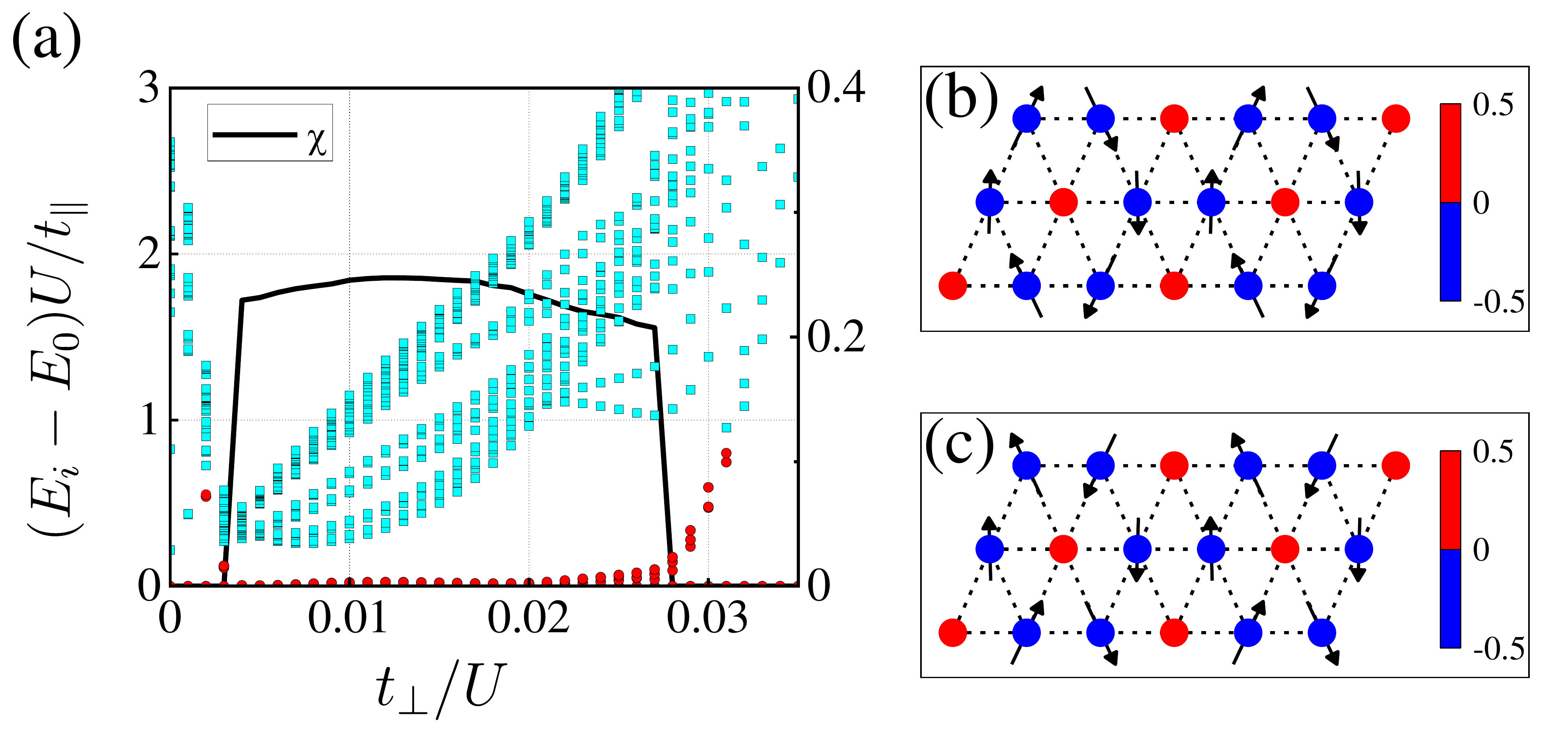}
\caption{(Color online) (a) Low energy spectra $E_i$ and scalar spin chirality $\chi$ as a function of $t_{\bot}/U$, obtained from exact diagonalizations, where the red dots denote the ground-state degeneracy and the green squares are for the excited states. ED calculations of real-space (b) Skyrmion and (c) anti-Skyrmion orbital textures~\cite{SM}, where the arrows represent the projection in the $xy$-plane of orbital polarization texture $\langle\boldsymbol{\mathcal S}_{{\bf r}} \rangle$, and the color denotes the $z$ component. Here, $t_{\parallel}/U \equiv 0.01$, $U=3U_1=3U_2$, and the lattice size $N_{\rm lat}=3\times6$. }
\label{ed_transition}
\end{figure}

To obtain more insights of the Skyrmion phases, we investigate ground-state degeneracy, scalar spin chirality $\chi$, and real-space orbital textures. As shown in Fig.~\ref{ed_transition}(a), we find that the Skyrmion phase is actually a gapped phase with a ground-state degeneracy and displays a finite scalar spin chirality order $\chi\neq0$~\cite{SM}, where both the degeneracy and $\chi$ disappear in other two Mott phases. We remark here that one only expects an approximate degeneracy
in simulations, due to finite-size effects. The ground-state degeneracy indicates the possibility of different types of orbital Skyrmion textures. As shown in Fig.~\ref{ed_transition}(b),(c), two different types of real-space orbital textures are resolved within ED~\cite{SM}. This prediction is consistent with BDMFT results [Fig.~\ref{fig_combine}(e)]. In addition, we also calculate orbital correlations between different lattice sites for the Skyrmion phase, and observe long-range correlations for the three components of the orbital polarization~\cite{SM}.

{\it Experimental detection.}
One key feature of the orbital Skyrmion lattice phase is the momentum structure shown in Fig.~\ref{transition_new} and~\ref{ed_phase}. Its spin analogue has been revealed by neutron scattering experiments~\cite{09Skyrmion}. The momentum structure of the orbital Skrymion lattice state can be probed by combining inter-orbital transition techniques~\cite{2014_Li_NC,2021_Zhou_OrbitalControl} and Bragg spectroscopy~\cite{1999_Ketterle_PRL}.
With the inter-orbital transition techniques~\cite{2014_Li_NC}, which has been demonstrated in experiments~\cite{2021_Zhou_OrbitalControl}, the orbital texture can be converted to density modulations, which maintain the same crystal structure.
The periodic density modulations can then be probed by the standard Bragg spectroscopy in cold atom experiments~\cite{1999_Ketterle_PRL}.

{\it Conclusion}.
We study cold atoms loaded in the $p$-orbital band of a triangular optical lattice, and find a chiral orbital Skyrmion lattice phase in a large part of the phase diagram.
This quantum state emerges due to natural anisotropic orbital-exchange interaction for $p$-orbital bosons, unlike the conventional Dzyaloshinskii-Moriya scenario. In this multi-orbital setting, the Skyrmion and anti-Skyrmion lattice states are exactly degenerate due to time-reversal symmetry, in contrast to the widely studied Skyrmion lattice states in spin systems.
The exotic orbital polarization texture of the orbital Skyrmion state can be probed by Bragg spectroscopy, a technique accessible to most cold atom experiments.

\textit{Acknowledgements.}
We acknowledge helpful discussions with W. Vincent Liu, Xiaoji Zhou, Yang Qi, Xuefeng Zhang, Ivana Vasi$\rm \acute{c}$, and Bo Liu.
This work is supported by National Natural Science Foundation of China (Grants No. 12074431, 11774428, 11934002),
National Program on Key Basic Research Project of China (Grant No. 2021YFA1400900), and Shanghai Science Foundation (Grants No.21QA1400500).
The numerical simulation  was carried out at National Supercomputer Center in Tianjin, and the calculations were performed on TianHe-1A.

\bibliography{references}

\bibliographystyle{apsrev4-1}

\begin{widetext}

\newpage

\begin{center}
\begin{center}
{\Large \bf Supplementary Material:  Emergent Orbital Skyrmion Lattice in a Triangular Atom Array}
\end{center}
\end{center}
\renewcommand{\theequation}{S\arabic{equation}}
\renewcommand{\thesection}{S-\arabic{section}}
\renewcommand{\thefigure}{S\arabic{figure}}
\renewcommand{\bibnumfmt}[1]{[S#1]}
\renewcommand{\citenumfont}[1]{S#1}
\setcounter{equation}{0}
\setcounter{figure}{0}

\section{Bosonic dynamical mean-field theory}
\subsection{ BDMFT details}
Dynamical mean-field theory (DMFT), an extension of the mean-field theory to a quantum version with local quantum fluctuations, is exact in the limit of infinite dimensionality where the self-energy is purely a local quantity. 
A major success of DMFT is the understanding of the Mott transition. In our paper, we utilize a bosonic dynamical mean-field theory (BDMFT) on the triangular lattice. BDMFT has been developed to provide a non-perturbative description of zero- and finite-temperature properties of the Bose-Hubbard model~\cite{Vollhardt, Hubener, Werner, Li2011, Li2012, Li2013, Liang15, Li2016, PhysRevLett.121.093401}, whose reliability of this approach has been compared against the quantum Monte-Carlo simulations~\cite{QMC_boson}.

BDMFT solves the lattice many-body problem by reducing the full lattices to a set of single-impurity problems. The impurity is embedded into a non-Markovian bath, which describes the interaction of the site with the rest of the lattice. The physics of the impurity lattice site $i$ is given by the local effective action
\begin{eqnarray}
\label{effective action}
\nonumber  \mathcal{S}^{(i)}_{\rm eff}&=&-\int d\tau d\tau^{ \prime} \sum_{\sigma \sigma^{\prime}} \boldsymbol{p}^{(i)}_{\sigma}\left(\tau \right)^\ast \boldsymbol{\mathcal{G}}^{(i)}_{0,\sigma \sigma^{\prime}} \left(\tau-\tau ^{ \prime} \right)^{-1} {\boldsymbol{p}^{(i)}_{\sigma^{\prime}}\left( \tau^{ \prime} \right)}^T+\int d \tau \left\{ \frac{U}{2}\sum_{\sigma}n^{(i)}_\sigma \left(\tau \right) \left[ n^{(i)}_\sigma \left(\tau \right)-1 \right] \right. \\ \nonumber
&+&\left. 2U_1n^{(i)}_x \left(\tau \right) n^{(i)}_y \left(\tau \right) + \frac{U_2}{2}\sum_{\sigma \neq\sigma^{\prime}}\left[p^{(i)}_{\sigma}\left(\tau \right)^{\ast} p^{(i)}_{\sigma}\left(\tau \right)^{\ast} p^{(i)}_{\sigma^\prime}\left(\tau \right) p^{(i)}_{\sigma^\prime}\left(\tau \right)\right] + \sum_{\left<ij \right>}t_{x} \left[p^{(i)}_{x}\left( \tau \right)^{\ast} \phi_{x,j}\left( \tau \right)+{\rm c.c.}\right]\right. \\
&+&\left. t_{y} \left[p^{(i)}_{y}\left( \tau \right)^{\ast} \phi_{y,j}\left( \tau \right)+{\rm c.c.} \right]+t_{xy} \left[p^{(i)}_{x}\left( \tau \right)^{\ast} \phi_{y,j}\left( \tau \right)+p^{(i)}_{y}\left( \tau \right)^{\ast} \phi_{x,j}\left( \tau \right)+{\rm c.c.}\right]\right\} ,
\end{eqnarray}
which explicitly depends on the site index $i$. In the effective action, $\tau$ is imaginary time, and $\sigma=x,y$ is the $p_{x}$ and $p_y$ orbits, respectively. $\boldsymbol{p}^{(i)}_{\sigma}\left(\tau \right)\equiv \left(p^{(i)}_{\sigma} \left(\tau \right),p^{(i)}_{\sigma}\left(\tau \right)^{\ast}\right)$ is the Nambu notation. $t_{x}$, $t_{y}$ and $t_{xy}$ are the hopping coefficients connected by $t_{\parallel}$ and $t_{\bot}$. Parameter $\phi_{\sigma,j}\left(\tau \right) \equiv\left<p_{\sigma,j} \left(\tau \right) \right>_0$, where $\left<...\right>_0$ denotes the expected value without the impurity site. The function $\boldsymbol{\mathcal{G}}^{(i)}_{0,\sigma \sigma^{\prime}}\left(\tau-\tau^{\prime}\right)$ is a local non-interacting propagator interpreted as a local dynamical Weiss mean-field which is a function of time instead of a single number. Thus dynamical Weiss mean-field takes local quantum fluctuations into account. The local self-energies $\Sigma^{(i)}_\sigma \left(i\omega_n\right)$ are obtained by solving the effective action.

In BDMFT, the lattice self-energy approximately coincides with the impurity self-energy. From the Dyson equation, the interacting lattice Green's function is obtained from
\begin{eqnarray}
\boldsymbol{G}_\sigma\left(i\omega_n\right)^{-1}=\boldsymbol{G}_{0,\sigma}\left(i\omega_n\right)^{-1}-\boldsymbol{\Sigma}_{\sigma}\left(i\omega_n\right),
\end{eqnarray}
where $\boldsymbol{G}_{0,\sigma}\left(i\omega_n\right)$ stands for the non-interacting Green's function
\begin{eqnarray}
\boldsymbol{G}_{0,\sigma}\left(i\omega_n\right)^{-1}=\left(\mu_\sigma+i\omega_n\right)\boldsymbol{1}-\boldsymbol{t},
\end{eqnarray}
with $\boldsymbol{1}$ being the unit matrix. The matrix elements $t_{ij}$ are hopping amplitudes for a given lattice. Note here that a boldface notation is used to denote a matrix with site-indexed elements $i$. By identifying the interacting local Green's functions with the diagonal elements of lattice Green's function $\boldsymbol{G}^{(i)}_\sigma\left(i\omega_n\right)=\left(\boldsymbol{G}_\sigma \left(i\omega_n\right)\right)_{ii}$,
we finally obtain the Weiss mean-field by the Dyson equation
\begin{eqnarray}
\label{lattice Green}
\boldsymbol{\mathcal{G}}^{(i)}_{0,\sigma}\left(i \omega_n \right)^{-1}=\boldsymbol{G}^{(i)}_\sigma \left( i\omega_n \right)^{-1}+\boldsymbol{\Sigma}^{(i)}_\sigma\left(i\omega_n\right),
\end{eqnarray}
which closes the self-consistency equation. The most difficult step of this loop is the solution of the effective action. In order to solve the effective action, the Anderson impurity model, which possesses the identical effective action with Eq.~(\ref{effective action}), is implemented. 
The self-consistency loop is solved as follows: starting from an initial choice for the Anderson impurity parameters and the superfluid order parameters, the Anderson impurity Hamiltonian is constructed in the Fock basis and diagonalized exactly to obtain the eigenstates and eigenenergies. The eigenstates and eigenenergies allow us to calculate the superfluid order parameter, the impurity Green's functions and self-energies. Then the lattice Green's functions are obtained. Subsequently, new Anderson impurity parameters are obtained, by comparing the new Weiss functions with the old ones. With these new Anderson impurity parameters, the procedure is iterated until converged.

For the results presented in this work, we consider the system with lattice sites up to $N_{\rm lat}=576$ and periodic boundary conditions. The maximum occupation number of the orbital for each normal bath is four to guarantee convergence in our simulations. In the calculations, random initial values are utilized for different lattice sites to break lattice-translational symmetry. We mainly focus on the unit filling case with $n_{p_x,i}+n_{p_y,i}=1$ in the Mott-insulating regime. The many-body phase diagram is shown in Fig. 1 in the main text. In the BDMFT calculation, we determine the phase boundaries by superfluid order $\phi_{x,y}$, stripe order $\Theta_{\rm stripe}$, and scalar spin chirality $\chi$, as shown in Fig. 2 in main text. Actually, the Mott-insulating phases can also be distinguished by real-space orbital texture $\langle \boldsymbol{\mathcal{S}_{r}}\rangle$. As shown in Fig.~\ref{measurement}, we clearly observe different real-space orbital textures for different Mott phases. The Skyrmion phase is shown in the first column~\cite{PhysRevB.91.224407,yu2018transformation,gao2020fractional}. The staggered phase in the second column is referred to be as a stripe Mott insulator (${\rm S_{MI}}$)~\cite{PhysRevLett.97.190406, PhysRevLett.97.190406, PhysRevLett.100.160403, PhysRevLett.100.200406}. The last phase is the ferro Mott-insulating phase (${\rm H_{MI}}$).

\begin{figure}[h]
\includegraphics[trim = 0mm 0mm 0mm 0mm, clip=true, width=0.6\textwidth]{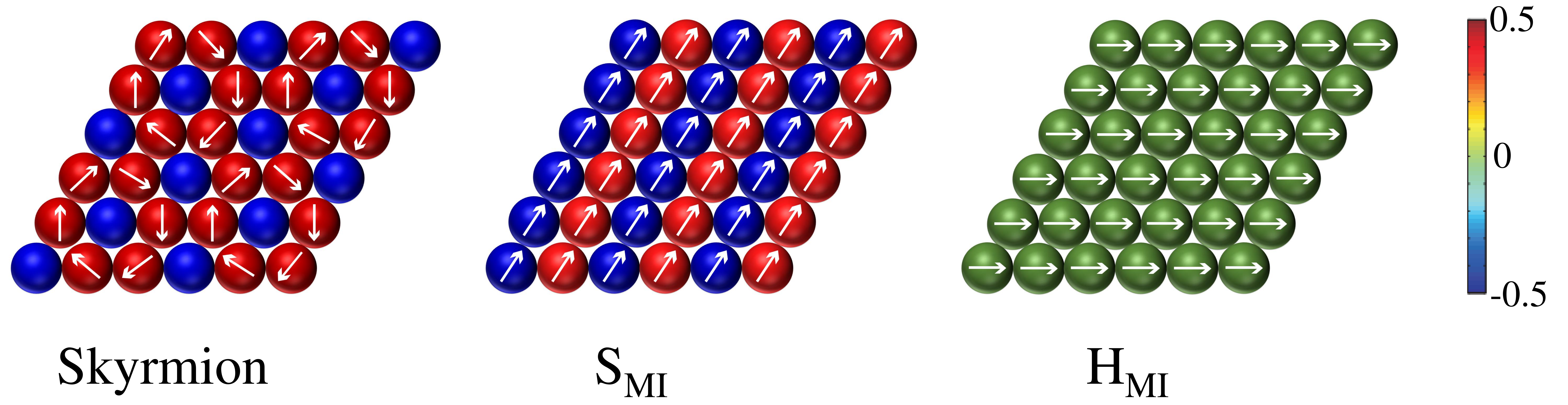}
\caption{(Color online) Real-space distributions of orbital polarization texture $\langle\boldsymbol{\mathcal{S}_{r}}\rangle$ for different Mott phases. The Skyrmion phase is shown in the first column. The ${\rm S_{MI}}$ and ${\rm H_{MI}}$ phases are given by the second and third columns, respectively. Here, the length of the arrows represents the amplitude of the local vectors in the $xy$-plane, and the color denotes the $z$-component.}
\label{measurement}
\end{figure}

We also study the robustness of Skyrmion texture against chemical potential. Our calculated filling-dependent phase diagrams are presented in Fig.~\ref{fig_mu}, as a function of chemical potential $\mu$ and hopping amplitudes (a) $t_{\parallel}=t_{\perp}$, and (b) $t_{\parallel}=5t_{\perp}$, respectively. We observe that there are three many-body quantum phases in the parameter regime studied here, including the Skyrmion, ${\rm S_{MI}}$, and ${\rm S_{SF}}$ phases. The orbital Skyrmion texture is robust and explores a large region of the phase diagrams in the lower hopping regime, indicating large opportunities for experimentally observing the many-body quantum phase. 
Our studies indicate that the Mott phase with Skyrmion texture is a general long-range order, stabilized in a large parameter regime.

\begin{figure}[h]
\includegraphics[trim = 0mm 0mm 0mm 0mm, clip=true, width=0.7\textwidth]{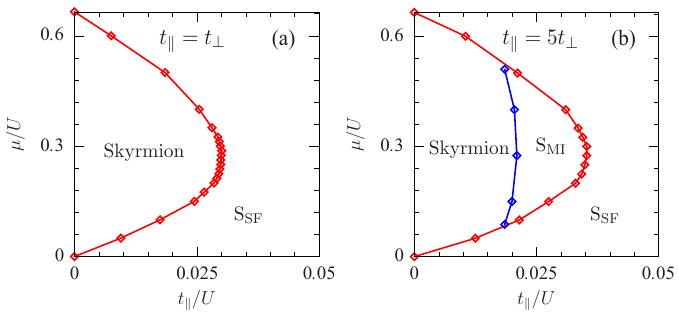}
\caption{(Color online) Filling-dependent phase diagrams of ultracold bosonic gases in $p$-orbital bands of a two-dimensional triangular lattice for different hopping amplitudes (a) $t_\parallel=t_\perp$ and (b) $t_\parallel=5t_\perp$, obtained via bosonic dynamical mean-field theory. The system supports two Mott-insulating phases with different types of orbital orders, including Skyrmion and stripe (${\rm S_{MI}}$) orbital textures, and the superfluid phase with stripe orbital angular momentum (${\rm S_{SF}}$). The interactions are $U=3U_1=3U_2$.}
\label{fig_mu}
\end{figure}

\section{EFFECTIVE ORBITAL-EXCHANGE MODEL}
Since the Mott-insulating phase is a state with suppressed number fluctuations, it is convenient to divide the Hilbert space with projection operator $\mathcal{P}$. For the Mott-insulator state with unit filling in the strong coupling limit $|t_{\parallel,\perp}| \ll U$, the $\mathcal{P}$ and $\mathcal{Q}=1-\mathcal{P}$, that project the Hilbert space into the Mott-insulating state and states with at least one site with double occupation, are introduced~\cite{auerbach2012interacting,essler2005one}. For Hamiltonian $H$, described by Eq.~(1) in the main text, we divided it into two parts $H=H_t+H_U$, that $H_t$ describes tunneling processes and $H_U$ describes interaction terms. The Schr{\"o}dinger equation reads
\begin{eqnarray}
\label{eq:1}
H\left|\psi\right>=(H_t+H_U)\left|\psi\right>=(H_t+H_U)(\mathcal{P}+\mathcal{Q})\left|\psi\right>=E\left|\psi\right>.
\end{eqnarray}
Multiplying the left side of Eq.~(\ref{eq:1}) by $\mathcal{P}$ and $\mathcal{Q}$ respectively results in
\begin{eqnarray}
\label{eq:2}
(\mathcal{P}H_t\mathcal{P}+\mathcal{P}H_t\mathcal{Q}+\mathcal{P}H_U\mathcal{P}+\mathcal{P}H_U\mathcal{Q})\left|\psi\right>=E\mathcal{P}\left|\psi\right>,\\
\label{eq:3}
(\mathcal{Q}H_t\mathcal{P}+\mathcal{Q}H_t\mathcal{Q}+\mathcal{Q}H_U\mathcal{P}+\mathcal{Q}H_U\mathcal{Q})\left|\psi\right>=E\mathcal{Q}\left|\psi\right>.
\end{eqnarray}
We remark here that $\mathcal{P}H_t\mathcal{P}$, $\mathcal{P}H_U\mathcal{P}$, $\mathcal{P}H_U\mathcal{Q}$ and $\mathcal{Q}H_U\mathcal{P}$ are zero. We can obtain a substitution from Eq.~(\ref{eq:3}) with the projection operator relation $\mathcal{Q}^2=\mathcal{Q}$,
\begin{eqnarray}
\label{eq:4}
\mathcal{Q}\left|\psi\right>=\frac{1}{E-\mathcal{Q}H_U\mathcal{Q}-\mathcal{Q}H_t\mathcal{Q}}\mathcal{Q}H_t\mathcal{P}\left|\psi\right>.
\end{eqnarray}
Inserting Eq.~(\ref{eq:4}) into Eq.~(\ref{eq:2}) with the projection operator relation, a revised equation for $\mathcal{P}\left|\psi\right>$ is given by
\begin{eqnarray}
\label{eq:5}
\left(\mathcal{P}H_t\mathcal{Q}\frac{1}{E-\mathcal{Q}H_U\mathcal{Q}-\mathcal{Q}H_t\mathcal{Q}}\mathcal{Q}H_t\mathcal{P}\right)\mathcal{P}\left|\psi\right>=E\mathcal{P}\left|\psi\right>.
\end{eqnarray}
Finally, the effective Hamiltonian in the deep Mott regime is given by~\cite{PhysRevLett.111.205302, PhysRevLett.121.015303}
\begin{eqnarray}
\label{eq:6}
\mathcal{H}_{\rm eff}=\mathcal{P}H_t\mathcal{Q}\frac{1}{E-\mathcal{Q}H_U\mathcal{Q}-\mathcal{Q}H_t\mathcal{Q}}\mathcal{Q}H_t\mathcal{P}.
\end{eqnarray}
Because $E\sim\frac{t^2}{U}$, we obtain $\frac{1}{E-\mathcal{Q}H_U\mathcal{Q}-\mathcal{Q}H_t\mathcal{Q}}\approx\frac{1}{-\mathcal{Q}H_U\mathcal{Q}-\mathcal{Q}H_t\mathcal{Q}}$. Using the expansion of $\frac{1}{-\mathcal{Q}H_U\mathcal{Q}-\mathcal{Q}H_t\mathcal{Q}}$ by $\frac{1}{A-B}=\frac{1}{A}\Sigma^{\infty}_{n=0}(B\frac{1}{A})^n$, with $A=-\mathcal{Q}H_U\mathcal{Q}$ and $B=\mathcal{Q}H_t\mathcal{Q}$, we obtain the effective Hamiltonian
\begin{eqnarray}
\label{eq:eff_ham}
\mathcal{H}_{\rm eff}&=&\mathcal{P}H_t\mathcal{Q}\frac{1}{-\mathcal{Q}H_U\mathcal{Q}}\sum^{\infty}_{n=0}\left(\mathcal{Q}H_t\mathcal{Q}\frac{1}{-\mathcal{Q}H_U\mathcal{Q}}\right)^n \mathcal{Q}H_t\mathcal{P}.
\end{eqnarray}

\subsection{Second-order effective Hamiltonian}
Firstly, we expand the effective Hamiltonian up to second-order $\mathcal{O} \left(t_{\parallel,\perp}^2/U \right)$, and Eq.~({\ref{eq:eff_ham}}) reads
\begin{eqnarray}
\label{eq:7}
\mathcal{H}_{\rm eff}&=&\mathcal{P}H_t\mathcal{Q}\frac{1}{-\mathcal{Q}H_U\mathcal{Q}}\mathcal{Q}H_t\mathcal{P}.
\end{eqnarray}
We focus on the unit-filling case with $n_{p_x,i}+n_{p_y,i}=1$. It is convenient to restrict the Hilbert space to a two-site problem. The basis spanning the subspace of states for projection operators $\mathcal{P}$ and $\mathcal{Q}$ is
\begin{eqnarray}
\mathcal{H}_{\mathcal{P}}&:&\left\{ \left|{p_x,p_x}\right>,\left|{p_x,p_y}\right>,\left|{p_y,p_x}\right>,\left|{p_y,p_y}\right> \right\}, \\
\mathcal{H}_{\mathcal{Q}}&:&\left\{ \left|{2p_x,0}\right>,\left|{p_x p_y,0}\right>,\left|{2p_y,0}\right> \right\},
\end{eqnarray}
where $\left|{p_{\alpha},p_{\beta}}\right>$ is a state with a $p_{\alpha}$-orbital atom at site $i$, and a $p_{\beta}$-orbital atom at site $j$. In the following, we replace the $p_{\alpha}$- and $p_{\beta}$-orbitals with $\uparrow$ and $\downarrow$. The subspace of projection operator is rewritten as
\begin{eqnarray}
\mathcal{H}_{\mathcal{P}}&:&\left\{ \left|{\uparrow,\uparrow}\right>,\left|{\uparrow,\downarrow}\right>,\left|{\downarrow,\uparrow}\right>,\left|{\downarrow,\downarrow}\right> \right\}, \\
\mathcal{H}_{\mathcal{Q}}&:& \left\{ \left|{\uparrow\uparrow,0}\right>,\left|{\downarrow\downarrow,0}\right>,\left|{\uparrow\downarrow,0}\right> \right\}.
\end{eqnarray}
Now, $\mathcal{Q}H_U\mathcal{Q}$ can be given in a matrix form
\begin{eqnarray}
\mathcal{Q}H_U\mathcal{Q}=\left( \begin{matrix}
U& \frac{U}{3}& 0\\
\frac{U}{3}& U& 0\\
0& 0& \frac{2U}{3}\\
\end{matrix} \right) ,
\end{eqnarray}
with the interactions $U=3U_1=3U_2$. The corresponding inverse of $\mathcal{Q}H_U\mathcal{Q}$ yields
\begin{eqnarray}
\left(\mathcal{Q}H_U\mathcal{Q}\right)^{-1}=\left( \begin{matrix}
\frac{9}{8U}& -\frac{3}{8U}& 0\\
-\frac{3}{8U}& \frac{9}{8U}& 0\\
0& 0& \frac{3}{2U}\\
\end{matrix} \right) .
\end{eqnarray}
Following Eq.~({\ref{eq:7}}), the effective Hamiltonian along the bond direction ${\bf {e}}_1$ is given by
\begin{eqnarray}
\label{eq:effe1}
H_{\rm eff}&=&-\underset{i}\sum\left[ \frac{9}{2U}t_{\parallel}^{2} n_{\uparrow ,i} n_{\uparrow ,i+{\bf e}_1}+\frac{9}{2U}t_{\bot}^{2} n_{\downarrow ,i} n_{\downarrow ,i+{\bf e}_1}+\frac{3}{2U}\left( t_{\parallel}^{2}+t_{\bot}^{2} \right) \left( n_{\uparrow ,i} n_{\downarrow ,i+{\bf e}_1}+n_{\downarrow ,i} n_{\uparrow ,i+{\bf e}_1} \right)\right.\nonumber \\
&+&\frac{3}{2U}t_{\parallel}t_{\bot} \left( p_{\downarrow ,i}^{\dagger} p_{\uparrow ,i} p_{\downarrow ,i+{\bf e}_1}^{\dagger} p_{\uparrow ,i+{\bf e}_1}+p_{\uparrow ,i}^{\dagger} p_{\downarrow ,i} p_{\uparrow ,i+{\bf e}_1}^{\dagger} p_{\downarrow ,i+{\bf e}_1} \right) \nonumber \\
&-&\left. \frac{3}{U}t_{\parallel}t_{\bot} \left( p_{\downarrow ,i}^{\dagger} p_{\uparrow ,i} p_{\uparrow ,i+{\bf e}_1}^{\dagger} p_{\downarrow ,i+{\bf e}_1}+p_{\uparrow ,i}^{\dagger} p_{\downarrow ,i} p_{\downarrow ,i+{\bf e}_1}^{\dagger} p_{\uparrow ,i+{\bf e}_1} \right) \right].
\end{eqnarray}
By introducing orbital polarization operators $\boldsymbol{\mathcal{S}}_i$
\begin{eqnarray}
\mathcal{S}^x_{i} &\equiv& \frac{1}{2} ({p}_{\uparrow,i} ^\dagger {p}_{\uparrow,i} - {p}_{\downarrow,i} ^\dagger {p}_{\downarrow, i}),\\
\mathcal{S}^y_{i} &\equiv& \frac{1}{2}({p}_{\uparrow,i} ^\dagger {p}_{\downarrow , i} + {p}_{\downarrow,i} ^\dagger {p}_{\uparrow , i}),\\
\mathcal{S}^z_{i} &\equiv& \frac{1}{2i}({p}_{\uparrow,i} ^\dagger {p}_{\downarrow, i} - {p}_{\downarrow,i} ^\dagger {p}_{\uparrow, i}),
\end{eqnarray}
we finally obtain
\begin{eqnarray}
\label{eq:e1}
H_{\rm eff}&=&\underset{i}\sum J_x\mathcal{S}^x_{i}\mathcal{S}^x_{i+{\bf e}_1}+J_y\mathcal{S}^y_{i}\mathcal{S}^y_{i+{\bf e}_1}+J_z\mathcal{S}^z_{i}\mathcal{S}^z_{i+{\bf e}_1},
\end{eqnarray}
where $J_x=-\frac{3}{2U}\left(t^2_{\parallel}+t^2_{\bot}\right)$, $J_y=\frac{3t_{\parallel}t_{\bot}}{U}$, and $J_z=\frac{9t_{\parallel}t_{\bot}}{U}$. We remark here that the parameter $J_x$ is ferro-orbital exchange, and $J_{y,z}$ is antiferro-orbital exchange.

The effective Hamiltonian along the bonds $\boldsymbol{e}_2$ and $\boldsymbol{e}_3$ can be obtained easily by rotating the coordinate of orbital polarization operators $\boldsymbol{\mathcal{S}}_{\bf r}$ ($\bf r$ utilized for lattice site in the main text). For a rotation along the bond ${\bf e}_m$ directing at angle $\theta_m$ with the $x$ axis, the $p$-orbital operator transforms as
\begin{eqnarray}
\tilde p_x&=&p_x {\rm cos}\theta_m +p_y {\rm sin} \theta_m , \\
\tilde p_y&=&-p_x{\rm sin}\theta_m  +p_y{\rm cos} \theta_m .
\end{eqnarray}
Accordingly, orbital polarization operator $\boldsymbol{\mathcal{S}}_{\bf r}$ becomes
\begin{eqnarray}
\label{Sx}
\tilde{\mathcal{S}}^x_{\bf r}\rightarrow \boldsymbol{ \mathcal{S}}_{{\bf r}}\cdot {\bf e}^x_{\theta_m}&=& {\rm cos} \left(2\theta_m \right) { \mathcal{S}}^x_{\bf r}+{\rm sin} \left(2\theta_m \right){ \mathcal{S}}^y_{\bf r} , \\
\label{Sy}
\tilde{\mathcal{S}}^y_{\bf r}\rightarrow\boldsymbol{ \mathcal{S}}_{{\bf r}}\cdot {\bf e}^y_{\theta_m} &=& -{\rm sin} \left(2 \theta_m \right) { \mathcal{S}}^x_{\bf r}+{\rm cos} \left(2 \theta_m \right){ \mathcal{S}}^y_{\bf r}, \\
\label{Sz}
\tilde{\mathcal{S}}^z_{\bf r}\rightarrow \boldsymbol{ \mathcal{S}}_{{\bf r}}\cdot {\bf e}^z_{\theta_m} &=& { \mathcal{S}}^z_{\bf r},
\end{eqnarray}
where ${\rm \theta_m}=0, \frac{1}{3}\pi$ and $\frac{2}{3}\pi$ for the bond directions ${\bf e}_1$, ${\bf e}_2$ and ${\bf e}_3$, respectively. We can introduce the rotation direction ${\bf e}_{\theta_m}$ for orbital polarization operator $\boldsymbol{\mathcal{S}}_{\bf r}$, {\rm i.e.}, ${\bf e}_{\theta_m}^x={\rm cos}\left(2 {\rm \theta_m} \right){\bf{e}}_x+{\rm sin}\left(2{\rm \theta_m}\right){\bf{e}}_y$, ${\bf e}_{\theta_m}^y=- {\rm sin}\left(2 {\rm \theta_m} \right){\bf{e}}_x+{\rm cos}\left(2{\rm \theta_m}\right){\bf{e}}_y$, and ${\bf e}_{\theta_m}^z={\bf e}_z$, which is used in the main text.

\subsection{Third-order effective Hamiltonian}
In the part, we expand the effective Hamiltonian, described by Eq.~({\ref{eq:eff_ham}}), up to third-order $\mathcal{O}(t_{\parallel,\perp}^3/U^2)$ terms, i.e.,
\begin{eqnarray}
\label{eq:8}
\mathcal{H}_{\rm  eff}&=&\mathcal{P}H_t\mathcal{Q}\frac{1}{\mathcal{Q}H_U\mathcal{Q}}\mathcal{Q}H_t\mathcal{Q}\frac{1}{\mathcal{Q}H_U\mathcal{Q}}\mathcal{Q}H_t\mathcal{P}.
\end{eqnarray}
Corresponding, the subspace of $\mathcal{H}_\mathcal{P}$ for a three-site problem with unit filling reads,
\begin{eqnarray}
\nonumber
\mathcal{H}_\mathcal{P}=&|&\uparrow,\uparrow,\uparrow \rangle, |\uparrow,\uparrow,\downarrow \rangle, |\uparrow,\downarrow,\uparrow \rangle, |\uparrow,\downarrow,\downarrow \rangle ,\\ \nonumber
&|& \downarrow,\uparrow,\uparrow \rangle, |\downarrow,\uparrow,\downarrow \rangle, |\downarrow,\downarrow,\uparrow \rangle, |\downarrow,\downarrow,\downarrow \rangle .
\end{eqnarray}
Following Eq.~(\ref{eq:8}), we can obtain a third-order effective Hamiltonian along the bond direction ${\bf e}_1$
\begin{eqnarray}
\label{eq:e2}
H_{\rm eff}&=&\underset{\Delta}\sum \left( J_{xxx} \mathcal{S}^x_{i}\mathcal{S}^x_{j}\mathcal{S}^x_k+J_{xyx}\mathcal{S}^x_{i}\mathcal{S}^y_{j}\mathcal{S}^x_k+J_{xyy}\mathcal{S}^x_{i}\mathcal{S}^y_{j}\mathcal{S}^y_k+J_{xzz}\mathcal{S}^x_i \tau^z_j \tau^z_k+J_{yxx}\mathcal{S}^y_i \mathcal{S}^x_j \mathcal{S}^x_k \right.\\ \nonumber
&+&\left.J_{yxy}\mathcal{S}^y_i \mathcal{S}^x_j \mathcal{S}^y_k+J_{yyx} \mathcal{S}^y_i \mathcal{S}^y_j \mathcal{S}^x_k+J_{yzz}\mathcal{S}^y_i \mathcal{S}^z_j \tau^z_k+J_{zxz}\mathcal{S}^z_i \mathcal{S}^x_j \mathcal{S}^z_k+J_{zyz}\mathcal{S}^z_i \mathcal{S}^y_j \mathcal{S}^z_k+J_{zzx}\mathcal{S}^z_i \mathcal{S}^z_j \mathcal{S}^x_k \right) \\ \nonumber
&+&\underset{i}\sum J_x^{\prime}\mathcal{S}^x_i \mathcal{S}^x_{i+{\bf e}_1}+J_y^{\prime}\mathcal{S}^y_i \mathcal{S}^{y}_{i+{\bf e}_1}+J_z^{\prime}\mathcal{S}^z_i \mathcal{S}^z_{i+{\bf e}_1}+J_{xy}^{\prime}\left( \mathcal{S}^x_i \mathcal{S}^y_{i+{\bf e}_1}-\mathcal{S}^y_i \mathcal{S}^x_{i+{\bf e}_1} \right),
\end{eqnarray}
where $\Delta$ denotes a triangular lattice with three sites $\left( i ,j ,k \right)=\left( i ,i+{\bf e}_1 , i+{\bf e}_2 \right)$. The parameters for the three-site terms are $J_{xxx}=\frac{27\left(t_{\parallel}+t_{\bot}\right)^3}{16U^2}$, $J_{yyx}=-\frac{27\left(3t^3_{\parallel}+t_{\parallel}^2 t_{\bot}+t_{\parallel} t_{\bot}^2+3t^3_{\bot}\right)}{16U^2}$, $J_{zzx}=-\frac{135\left(3t^3_{\parallel}+t_{\parallel}^2 t_{\bot}+t_{\parallel} t_{\bot}^2+3t^3_{\bot}\right)}{16U^2}$, $J_{xyx}=-J_{yxx}= -\frac{27\sqrt{3}\left(t^3_{\parallel}-t^2_{\parallel}t_{\bot}-t_{\parallel}t^2_{\bot}+t^3_{\bot}\right)}{16U^2}$, $J_{xyy}=J_{yxy}=-\frac{27t_{\parallel}t_{\bot}\left(t_{\parallel}+t_{\bot}\right)}{4U^2}$, $J_{xzz}=J_{zxz}=\frac{135\left(3t^3_{\parallel}+t_{\parallel}^2 t_{\bot}+t_{\parallel} t_{\bot}^2+3t^3_{\bot}\right)}{32U^2}$, and $J_{zyz}=-J_{yzz}=-\frac{135\sqrt{3}\left(3t^3_{\parallel}+t_{\parallel}^2 t_{\bot}+t_{\parallel} t_{\bot}^2+3t^3_{\bot}\right)}{32U^2}$, respectively, and the parameters for the two-site terms are
$J^{\prime}_x=-\frac{99\left(t^3_{\parallel}+9t^2_{\parallel}t_{\bot}-9t_{\parallel}t^2_{\bot}-t^3_{\bot}\right)}{64U^2}$,
$J^{\prime}_y=\frac{99\left(3t^3_{\parallel}-5t^2_{\parallel}t^{\bot}+5t_{\parallel}t^2_{\bot}-3t^3_{\bot}\right)}{64U^2}$, $J^{\prime}_z=\frac{279\left(3t^3_{\parallel}-5t^2_{\parallel}t^{\bot}+5t_{\parallel}t^2_{\bot}-3t^3_{\bot}\right)}{64U^2}$, and $J_{xy}^{\prime}=\frac{99\sqrt{3}\left(t^3_{\parallel}+t^2_{\parallel}t_{\bot}-t_{\parallel}t^2_{\bot}-t^3_{\bot}\right)}{64U^2}$. The final effective Hamiltonian is written as
\begin{eqnarray}
\label{eq:e3}
H_{\rm eff} &=&\sum_{{\bf r},m}\left(J_x+J^{\prime}_x\right) \left[\boldsymbol{\mathcal{S}}_{{\bf r}}\cdot {\bf e}^x_{\theta_m}\right]\left[\boldsymbol{\mathcal{S}}_{{\bf r}+{\bf{e}}_m}\cdot {\bf e}^x_{\theta_m}\right]
+\left(J_y+J^{\prime}_y\right)\left[\boldsymbol{\mathcal{S}}_{{\bf r}}\cdot {\bf e}^y_{\theta_m}\right]\left[\boldsymbol{\mathcal{S}}_{{\bf r}+{\bf{e}}_m}\cdot {\bf e}^y_{\theta_m}\right] \nonumber \\
&+&\left(J_z+J^{\prime}_z\right)\left[\boldsymbol{\mathcal{S}}_{{\bf r}}\cdot {\bf e}^z_{\theta_m}\right]\left[\boldsymbol{\mathcal{S}}_{{\bf r}+{\bf{e}}_m} \cdot {\bf e}^z_{\theta_m}\right]\nonumber \\
&+&J_{xy}^{\prime}\left\{\left[\boldsymbol{ \mathcal{S}}_{\bf r}\cdot {\bf e}^x_{\theta_m}\right]\left[\boldsymbol{ \mathcal{S}}_{{\bf r}+\bf{e}_m}\cdot{\bf e}^y_{\theta_m} \right] - \left[\boldsymbol{ \mathcal{S}}_{\bf r}\cdot {\bf e}^y_{\theta_m}\right]\left[\boldsymbol{ \mathcal{S}}_{{\bf r}+\bf{e}_m}\cdot{\bf e}^x_{\theta_m} \right]\right\}\nonumber \\
&+&\sum_{{\bf r},u,v,w}J^{\prime}_{uvw}\left[\boldsymbol{\mathcal{S}}_{\bf r}\cdot {\bf e}^u_{\theta_1}\right]\left[ \boldsymbol{\mathcal{S}}_{{\bf r}+{\bf e}_1} \cdot {\bf e}^v_{\theta_1}\right]\left[\boldsymbol{\mathcal{S}}_{{\bf r}+{\bf e}_2}\cdot {\bf e}^w_{\theta_1}\right] ,
\end{eqnarray}
where the $J^\prime_{xy}$ term is the $\it z$-component of the cross product. We remake here that the cross-product term plays tiny role in our simulations, which does not influence the orbital structures but only shifts the phase boundary slightly. The orbital Skyrmion texture is actually a result of anisotropic orbital-exchange interactions.

\begin{figure}[h]
\includegraphics[trim = 0mm 0mm 0mm 0mm, clip=true, width=0.475\textwidth]{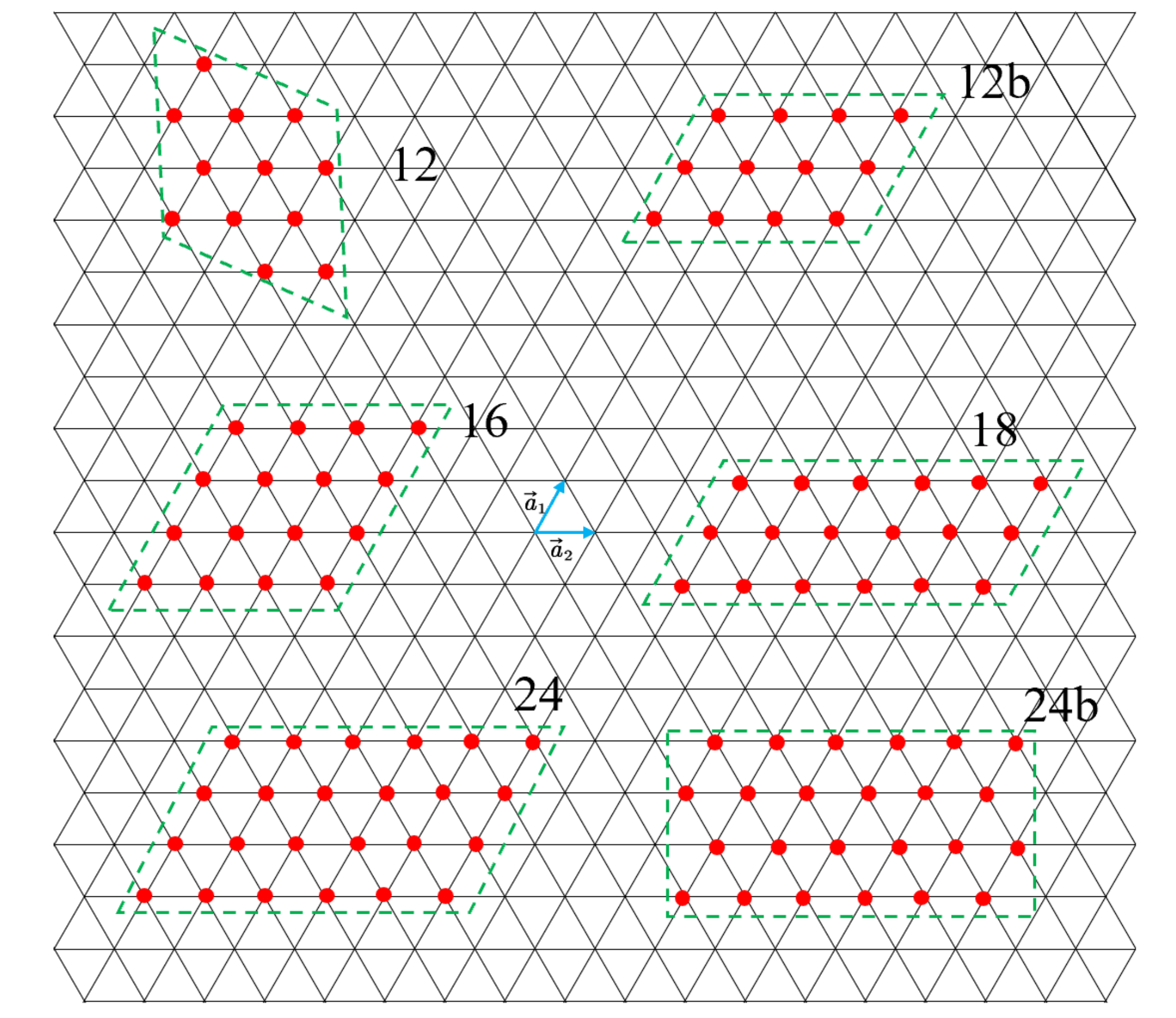}
\caption{(Color online) Clusters used in the ED calculations. $\vec{a}_1=\left(a/2,\sqrt{3}a/2 \right)$ and $\vec{a}_2=\left(a,0 \right)$ are the primitive vectors of the triangular lattice. The clusters 12, 16, 24, and 24b contain the M points in the reciprocal space, and the lattices 12 and 24b contain the K momentum points. We mainly utilize the cluster 24b in our ED calculation to contain both M and K points.  }
\label{cluster}
\end{figure}

\section{EXACT DIAGONALIZATION}
\subsection{Clusters in ED calculation}
In this paper, we mainly utilize Lanczos exact diagonalization to solve the effective orbital-exchange model, and obtain phase diagrams and low-energy spectra. The ED calculations are under periodic boundary conditions, and the largest system size considered here is $24$ lattice sites. The clusters used in our ED calculations are shown in Fig.~\ref{cluster} for different lattice structures, denoted as $12$, 12b, $16$, $18$, $24$, and 24b, respectively. Among these clusters, the lattices $12$ and 24b contain the K momentum points in the reciprocal space, and the clusters $12$, $16$, $24$, and 24b contain the M momentum points in the reciprocal space. The cluster, mainly used in our ED calculations, is 24b.
\begin{figure}[h]
\includegraphics[trim = 0mm 0mm 0mm 0mm, clip=true, width=0.6\textwidth]{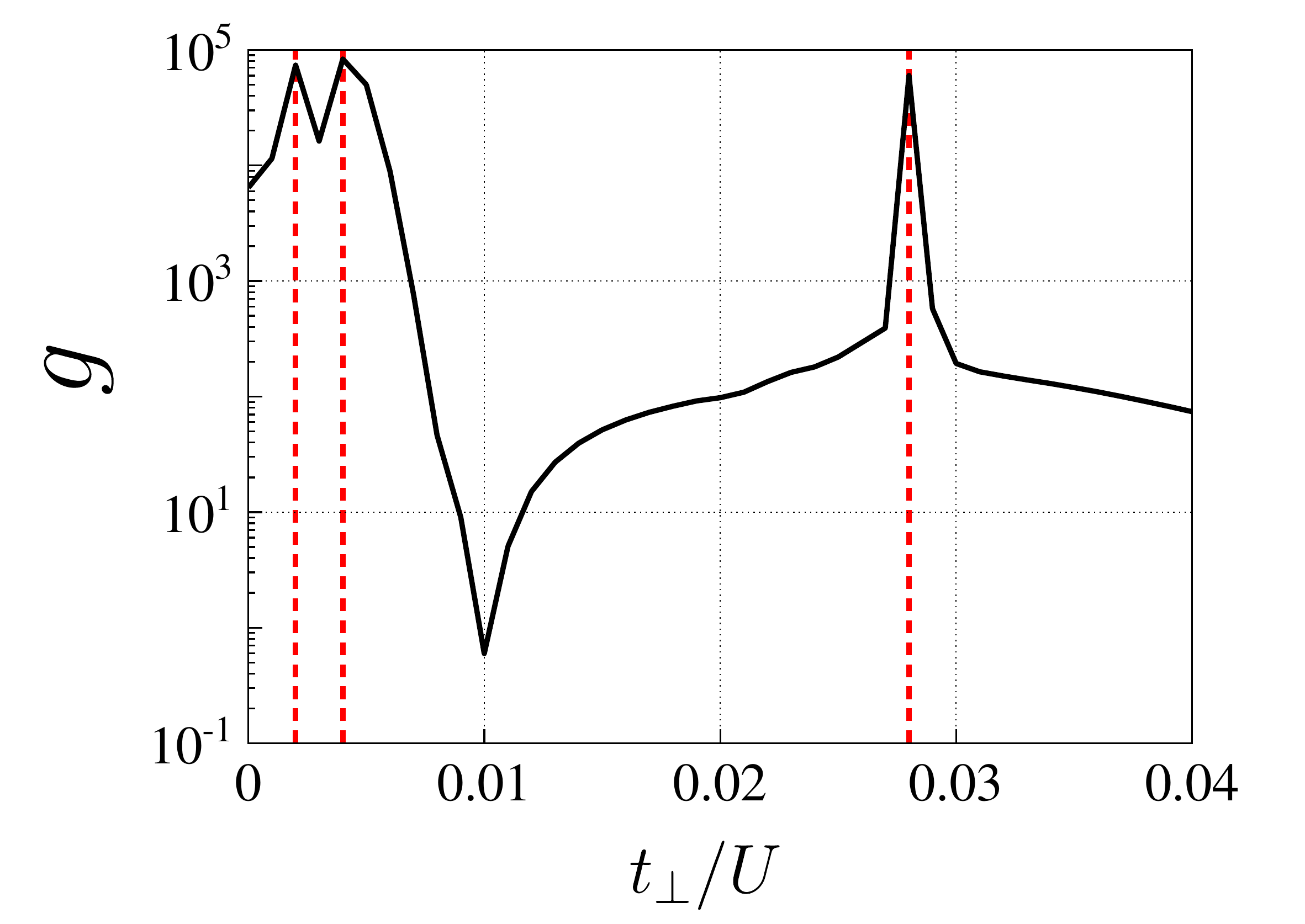}
\caption{(Color online) ED calculation of the fidelity metric $g$ on a lattice 24b for $t_{\parallel}/U \equiv 0.01$. The red-dotted lines are the phase-transition points.}
\label{fidelity}
\end{figure}
\subsection{Details of phase diagram within effective Hamiltonian}
To determine phase boundaries of the effective orbital-exchange Hamiltonian, we also calculate the fidelity metric $g$~\cite{gu2010fidelity, PhysRevE.74.031123, PhysRevB.82.115125}. Supposing $|\psi_0 \left(\lambda \right) \rangle$ being the ground state of $H\left(\lambda \right)$, and $|\psi_0 \left(\lambda +\delta \lambda \right)\rangle$ the ground state of $H\left(\lambda+\delta \lambda \right)$, the fidelity metric $g$ is given by
\begin{eqnarray}
g\left(\lambda,\delta \lambda \right)\equiv \frac{2}{N} \frac{1-|\langle \psi_0\left(\lambda \right)|\psi_0\left(\lambda+\delta \lambda \right)\rangle|}{\left(\delta \lambda \right)^2},
\end{eqnarray}
where $N$ is the number of lattice sites, and $\delta \lambda=0.001$ for $\lambda=t_{\bot}/U$ or $ t_{\parallel}/U$. The result for the cluster 24b is shown in Fig.~\ref{fidelity}, where diverging peaks are observed around the phase-transition points.
In the main text, we have utilized both fidelity metric and spin structure factor to determine phase boundaries, as shown in Fig. 2. To provide more information about the Skyrmion phase, real-space orbital correlations are shown in the Fig.~\ref{correlation} for the cluster 24b.

\begin{figure}[h]
\includegraphics[trim = 0mm 0mm 0mm 0mm, clip=true, width=\textwidth]{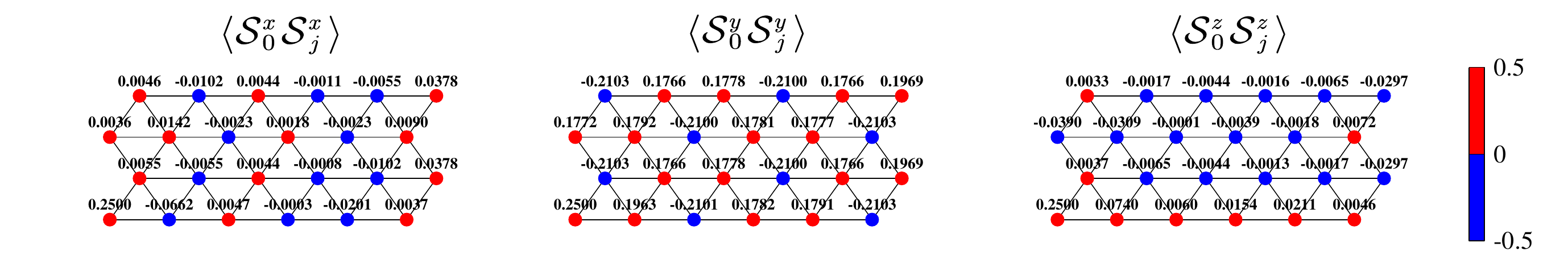}
\caption{(Color online) Real-space orbital correlations of the Skyrmion phase. The red dots indicate a positive correlation between sites 0 and $j$, and the blue dots denote a negative correlation. The lattice structure is 24b, $t_{\parallel}/U=0.01$, and $t_{\bot}/U=0.026$.}
\label{correlation}
\end{figure}
Generally, one cannot obtain the real-space orbital polarization $\langle \mathcal{S}_i^\alpha\rangle$ within ED calculation directly. Instead, we need to use the correlation matrix $\left[M^{\alpha}\right]_{i,j}=\langle \mathcal{S}^{\alpha}_i \mathcal{S}^{\alpha}_j \rangle$ to construct orbital textures, where $\alpha=x,y,z$, and $i$ and $j$ denote lattice sites. We extract the orbital polarization pattern in the ED calculation by diagonalizing the three correlation matrices $\left[ M^{\alpha = x,y,z}\right]_{i,j}$.
The maximal eigenvalues and the corresponding eigenvectors are denoted as $\lambda^\alpha$ and $\boldsymbol{u}^\alpha$, respectively.
The orbital polarization $\langle \boldsymbol{\mathcal{S}_i} \rangle$ at each site is then obtained by rescaling of $\sqrt{\lambda^\alpha}\boldsymbol{u}^\alpha$~\cite{li2013topological}. The real-space orbital textures of ED are shown in Fig.~4(b),(c) in the main text. After obtaining polarization $\langle \boldsymbol{\mathcal{S}_i} \rangle$, scalar spin chirality can be constructed correspondingly, as shown in Fig.~4(a) in the main text.

\end{widetext}

\end{document}